\documentclass[
  onecolumn,
  amsmath,
  amssymb
]{revtex4-2}

\usepackage{graphicx}
\usepackage{dcolumn}
\usepackage{epstopdf}
\usepackage{bm}
\usepackage{hyperref}
\hypersetup{
    colorlinks=true,   
    citecolor=blue,    
    linkcolor=blue,    
    urlcolor=blue      
}
\usepackage{booktabs}
\usepackage{xcolor}
\usepackage{placeins}
\usepackage[mathlines]{lineno}
\newcommand{\sgn}{Sc$_{x}$Ga$_{1\text{-}x}$N}

\begin{document}

\title{\textbf{Local coordination, structural softening, and polarization-switching energetics in Sc-alloyed GaN}}

\author{Shailesh Kalal$^{1,\#}$}
\email{shailesh.kalal@liu.se}
\author{Gueorgui Kostov Gueorguiev$^{2,\#}$}
\author{Martin Magnuson$^1$}
\author{Edward Ferraz de Almeida Junior$^3$}
\author{Rohini Sanikop$^1$}
\author{Sagar Jathar$^4$}
\author{Per Sandström$^1$}
\author{Ray-Hua Horng$^{1,5}$}
\author{Jens Birch$^1$}
\author{Per Eklund$^{1,4}$}
\author{Ching-Lien Hsiao$^1$}
\email{ching-lien.hsiao@liu.se}

\thanks{$^\#$These authors contributed equally.}

\affiliation{$^1$Thin Film Physics Division, Department of Physics, Chemistry and Biology (IFM), SE 58183 Linköping University, Linköping, Sweden}
\affiliation{$^2$LiU-UFBA Research and Education Network, Linköping University, SE 58183 Linköping, Sweden}
\affiliation{$^3$Center for Exact Sciences and Technologies, Federal University of Western Bahia, Bertioga Street, 892, Morada Nobre I, Barreiras, 47810-059, Brazil}
\affiliation{$^4$Inorganic Chemistry, Department of Chemistry - Ångström Laboratory, Uppsala University, Box 538, SE-751 21 Uppsala, Sweden}
\affiliation{$^5$Institute of Electronics, National Yang Ming Chiao Tung University, Hsinchu 30010, Taiwan}


\begin{abstract}

Sc-alloyed wurtzite nitrides exhibit strongly tunable electromechanical and ferroelectric properties, yet the relationship between their lattice evolution and local bonding environment remains insufficiently established, particularly in \sgn{}. Here, we investigate the structural and local bonding evolution of \sgn~across the dilute-to-intermediate composition range by combining X-ray diffraction, Sc $K$-edge X-ray absorption near edge structure (XANES), extended X-ray absorption fine structure (EXAFS)  and first-principles calculations. Sc incorporation produces an anisotropic lattice expansion and a progressive reduction in the $c/a$ ratio. XANES and EXAFS reveal a concurrent modification of the local environment around Sc, with the effective Sc-N coordination number increasing from $4.1(4)$ to $4.5(2)$ and the average Sc-N bond length increasing from $2.045(7)$ to $2.081(8)$~\AA\ over $x=0.06$-$0.26$. The local response is accompanied by a reduction in the Sc $K$-edge pre-edge intensity, consistent with a gradual reduction of the local tetrahedral asymmetry. First-principles calculations show that these structural changes are associated with site-selective distortions around Sc and progressive flattening of the structural energy landscape. The calculated intrinsic polarization-switching barrier decreases from $24.2$ to $19.0$~meV/\AA$^3$ with increasing Sc content, while the calculated piezoelectric stress coefficient ($e_{33}$) increases from $0.82$ to $1.66$~C/m$^2$ and elastic constant ($C_{33}$) decreases from $380$ to $227$~GPa, resulting in an increase of piezoelectric strain coefficient ($d_{33}$) from $2.99$ to $12.46$~pC/N. These results show that Sc incorporation progressively modifies the local coordination environment and structural energetics of ScGaN while the long-range wurtzite structure remains preserved over the investigated composition range.

\end{abstract}

\maketitle
\section{Introduction}

Wurtzite III-nitride semiconductors combine wide-bandgap electronic functionality with strong spontaneous and piezoelectric polarization, enabling applications ranging from polarization-induced two-dimensional carrier gases and polarization doping to high-frequency, power-electronic, optoelectronic, and electromechanical devices \cite{Bernardini1997,2002_Ambacher}. The demonstration of switchable polarization in Sc-alloyed III-nitrides has further expanded this materials family from conventional polar semiconductors to ferroelectric semiconductors \cite{2009_Akiyama,2019_Fichtner}. This combination of semiconductor and ferroelectric functionality is particularly attractive for controlling interface charge, carrier density, and electromechanical coupling in nitride-based devices. \cite{Wang_2021,Xie2025,Zeng2026nitride} Realizing these opportunities, however requires a microscopic understanding of how alloy chemistry modifies the local atomic structure and consequently, the polarization-switching and electromechanical responses. 

The unusual response of Sc-alloyed wurtzite nitrides originates from a coordination mismatch. Whereas Ga and Al favor fourfold tetrahedral coordination in the wurtzite structure, Sc favors higher coordination, approaching the sixfold environment of rocksalt ScN \cite{Tasnadi2010PRL,ZHU2026}. Incorporation of Sc therefore introduces local competition within the tetrahedral wurtzite framework and modifies the structural parameters that describe its polar geometry, including the internal parameter $u$ and axial ratio $c/a$. In ScAlN, this coordination competition has been associated with a flattened structural energy landscape, enhanced piezoelectric response, and reduced polarization-switching barriers \cite{Tasnadi2010PRL,Tholander2013PRB,2019_Fichtner,Calderon2023Science,Lee2024SciAdv}. Recent atomistic studies further indicate that polarization reversal can involve heterogeneous, locally distorted configurations rather than a spatially uniform structural pathway \cite{Calderon2023Science,Lee2024SciAdv,Paillard2026,Ye2025NatCommun}. These results have established local structural flexibility as an important ingredient in the response of Sc-alloyed wurtzite ferroelectrics. However, most of this microscopic understanding has been developed for ScAlN, leaving open the question of whether the same structural mechanism applies to other Sc-alloyed III-nitrides.

In contrast to ScAlN, ScGaN introduces Sc into the technologically mature GaN lattice and therefore combines the coordination preference of Sc with a substantially larger wurtzite lattice and a different cation-anion bonding environment. ScGaN retains the semiconductor characteristics of the III-nitride family while exhibiting pronounced piezoelectric and ferroelectric responses. Ferroelectric switching has been demonstrated in both epitaxial and sputter-deposited ScGaN, with remanent polarization approaching $120~\mu$C~cm$^{-2}$ \cite{Wang_2021,Uheara201_APL,Yang_2025}. Increasing Sc concentration also reduces the coercive field, and phase-pure monocrystalline \sgn~ has recently been reported up to $x=0.48$, with coercive fields approaching $1.2$~MV~cm$^{-1}$ \cite{Yang_2025}. ScGaN has furthermore been reported to exhibit a lower coercive field than ScAlN at comparable remanent polarization \cite{Uehara_2022}. Comparative first-principles studies indicate that the composition-driven structural transformation in ScAlN is dominated by changes along the polar direction, whereas ScGaN exhibits appreciable structural changes in both the in-plane and out-of-plane directions \cite{Pike2025}. This distinction is consistent with earlier experimental observations of anisotropic lattice expansion in dilute ScGaN, where the in-plane lattice parameter was found to increase substantially more than the out-of-plane parameter \cite{2004_Constantin}. The larger lattice and different bonding environment of GaN therefore provide an additional structural degree of freedom for accommodating the higher-coordination preference of Sc. This raises a fundamental question: how does the Sc-centered local environment reorganize when Sc is incorporated into GaN, and how does this local response relate to the macroscopic lattice evolution and polarization-switching energetics? Addressing this question is particularly relevant because average lattice parameters alone cannot determine whether the structural response is distributed throughout the alloy or localized around the Sc atoms.

Experimental information on the local structure of ScGaN remains limited. Previous X-ray absorption (XAS) studies of dilute ScGaN, restricted to $x\leq0.059$, showed that Sc occupies a distorted tetrahedral environment characterized by an increased local internal parameter $u$ \cite{Knoll2014JPCM}. Whether this dilute-limit configuration persists at higher Sc concentrations or progressively evolves toward higher-coordination configurations has not been established experimentally. In particular, it remains unclear whether the composition-dependent reduction in $c/a$ reflects a homogeneous deformation of the wurtzite lattice or a site-selective rearrangement of the Sc-centered coordination environment. Establishing this distinction requires combining average structural probes with element-specific measurements of the local Sc environment. A corresponding experimental connection between Sc-centered coordination, anisotropic lattice evolution, and the energetics of polarization reversal is presently lacking for ScGaN.

In this work, we combine composition-dependent X-ray diffraction (XRD), Sc $K$-edge XAS, including X-ray absorption near edge structure (XANES) and extended X-ray absorption near edge structure (EXAFS), with first-principles calculations to investigate the structural evolution of wurtzite \sgn~across the dilute-to-intermediate composition range ($x\leq0.26$). XRD is used to determine the average lattice evolution, while XANES and EXAFS provide element-specific information on the local symmetry, bond lengths, and coordination environment around Sc. These measurements are complemented by first-principles calculations of the local structural parameters, structural energy landscapes, electronic states, electromechanical response, and polarization-switching pathways. This combined approach allows the local and average structural responses of ScGaN to be examined within a common framework and provides a basis for relating coordination evolution to structural softening and polarization-switching energetics.

\FloatBarrier

\section{Lattice evolution and anisotropic structural flattening}

The composition of the \sgn~films grown on Al$_2$O$_3$ (0001) was determined by Rutherford backscattering spectrometry (RBS) and time-of-flight elastic recoil detection analysis (ToF-ERDA). Representative spectra are shown in Fig.~S1 of the Supporting Information (SI), and the extracted Sc molar fractions are $x=0.00$, $0.06$, $0.15$, and $0.26$ (Table~S1). All films exhibit uniform depth profiles and impurity concentrations below the detection limit ($\sim$0.5 at.\%), confirming high compositional homogeneity across the alloy series.

\begin{figure}
\centering
\includegraphics[width=1\textwidth]{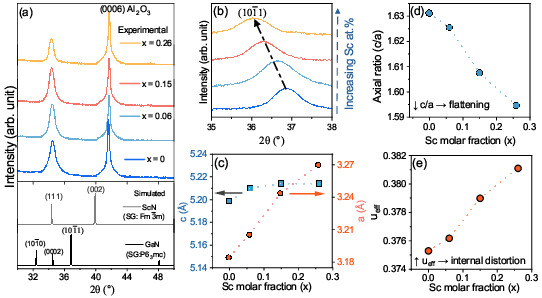}
\caption{(a) XRD $\omega$-2$\theta$ scans of \sgn~thin film along with reference diffraction patterns for wurtzite GaN ($P6_3mc$) and rocksalt ScN ($Fm\bar{3}m$). (b) Enlarged view of the asymmetric (10$\bar{1}$1) reflection measured using $\chi$ $\approx$ 61.9$^\circ$. (c) $a$ and $c$ lattice parameters as a function of Sc composition. (d) Reduction of the axial ratio $c/a$ with increasing Sc concentration. (e) Effective internal parameter $u_{\mathrm{eff}}$, estimated using the equal-bond approximation.}
\label{fig:fig1XRD}
\end{figure}

The global structural evolution of \sgn~was investigated by XRD. Figure~\ref{fig:fig1XRD}(a) shows the $\omega$-$2\theta$ scans together with the simulated diffraction pattern of wurtzite GaN ($P6_3mc$) and rocksalt type ScN ($Fm\bar{3}m$). All Sc-containing films up to $x=0.26$ exhibit intense (0002) reflections characteristic of phase-pure, $c$-axis-oriented wurtzite growth, with no detectable secondary phases within the resolution of the measurement~\cite{Hsiao2026APL,PINGEN2025}. With increasing Sc content, the (0002) reflection shifts systematically toward lower $2\theta$ and becomes broader, indicating an increase in the average lattice spacing accompanied by enhanced structural disorder and microstrain. A corresponding composition-dependent evolution is observed for the asymmetric (10$\bar{1}$1) reflection (Fig.~\ref{fig:fig1XRD}(b)).

The lattice parameters extracted from the symmetric and asymmetric reflections reveal a strongly anisotropic response (Fig.~\ref{fig:fig1XRD}(c)). The in-plane lattice parameter $a$ increases substantially with Sc incorporation, whereas the out-of-plane parameter $c$ exhibits a comparatively weaker variation. Consequently, the axial ratio $c/a$ decreases monotonically from the GaN value with increasing Sc concentration (Fig.~\ref{fig:fig1XRD}(d)), resulting in a progressive flattening of the wurtzite lattice. Such anisotropic lattice expansion is consistent with previous structural studies of Sc-alloyed AlN and has been associated with the local distortions introduced by Sc substitution~\cite{Lee2024SciAdv,Chen2025PRM}. Furthermore, to quantify the associated deviation from the ideal wurtzite geometry, we estimate an effective internal coordinate using the equal-bond approximation,
\begin{equation}
u_{\mathrm{eff}}=\frac{a^2}{3c^2}+\frac{1}{4}.
\end{equation}
Although $u_{\mathrm{eff}}$ does not represent the microscopic internal coordinate of an individual cation site, it provides a useful geometric descriptor of the average distortion of the wurtzite tetrahedral framework~\cite{SCHULZ1977,Bernardini1997}. As shown in Fig.~\ref{fig:fig1XRD}(e), $u_{\mathrm{eff}}$ increases systematically with Sc concentration, while $c/a$ decreases. The simultaneous evolution of these two structural parameters indicates a progressive departure from the ideal wurtzite geometry, with increasing flattening of the tetrahedral framework. Importantly, the XRD-derived parameters represents an average structural parameters and do not by resolve whether the distortion is uniformly distributed throughout the alloy or localized around Sc atoms. The systematic increase in $u_{\mathrm{eff}}$ and reduction in $c/a$, however, establish that Sc incorporation progressively modifies the geometry of the wurtzite lattice. The origin of this distortion is addressed next through the local structural and coordination analysis.

\FloatBarrier

\section{Local symmetry evolution and orbital hybridization}

\begin{figure}
\centering
\includegraphics[width=1\linewidth]{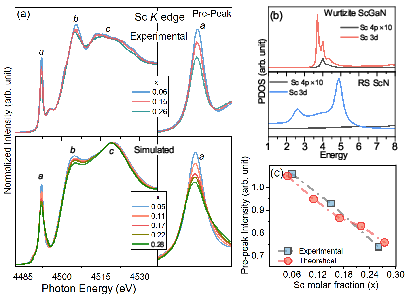}
\caption{(a) Experimental and simulated Sc $K$-edge XANES spectra for \sgn, showing the systematic evolution of features $a$, $b$, and $c$ with increasing Sc concentration. The inset highlights the pre-edge feature ($a$). (b) Projected density of states for wurtzite ScGaN and rocksalt ScN. (c) Experimental and calculated pre-edge intensities as a function of Sc molar fraction.}

\label{fig:fig2XANES}
\end{figure}

To examine how Sc incorporation alters the local bonding environment, we investigated the Sc $K$-edge XANES and compared the experimental spectra with first-principles simulated. The normalized experimental spectra are shown in Fig.~\ref{fig:fig2XANES}(a). Herein, all compositions exhibit a pronounced pre-edge feature ($a$) followed by broader near-edge peaks ($b$ and $c$). The calculated partial density of states (PDOS) shown in Fig.~\ref{fig:fig2XANES} (b) supports that the pre-edge originates primarily from nominally dipole-forbidden $1s\rightarrow3d$ transitions that gain dipole intensity through Sc $3d$-$4p$ hybridization in a locally non-centrosymmetric environment. It makes this feature extremely sensitive to the local symmetry and bonding configuration~\cite{deGroot_2009,KUMAR2025182507}. Interestingly, the pre-edge intensity decreases progressively with increasing Sc concentration (see enlarged view of feature \textit{a} in Fig.~\ref{fig:fig2XANES} (a)), indicating a gradual modification of the local symmetry and Sc $3d$-$4p$ hybridization. In conjunction with the XRD-derived decrease in axial ratio and increase in $u_{\mathrm{eff}}$, this evolution is consistent with a progressive departure of the Sc-centered environment from the ideal tetrahedral geometry. Importantly, a distinct pre-edge feature remains observable at $x=0.26$, indicating that the local Sc environment retains non-centrosymmetric character within the investigated composition range and does not approach a fully centrosymmetric octahedral configuration~\cite{ZHU2026}.

Furthermore, the simulated XANES spectra reproduce the principal composition-dependent evolution of the pre-edge feature (Fig.~\ref{fig:fig2XANES} (a)), suggesting its association with changes in the Sc-centered bonding environment. Thus, the experimental \sgn~spectra indicate a continuous evolution of the Sc-centered environment away from an ideal tetrahedral configuration while the long-range wurtzite structure remains preserved. The XANES spectra for reference rocksalt (RS) ScN films were also experimentally measured and simulated as given in Fig.~S2 (a) of SI, which further illustrates the sensitivity of the pre-edge to local symmetry and orbital hybridization. PDOS shown in Fig.~\ref{fig:fig2XANES} (b) confirms reduced $3d$-$4p$ mixing in centrosymmetric rocksalt ScN, resulting in substantially weaker pre-edge features (Fig.~S2 in SI). Thus, the experimental \sgn~spectra indicate a continuous evolution of the local Sc environment toward a tetrahedrally distorted configuration while the long-range wurtzite structure remains preserved. Furthermore, the Ga $K$-edge spectra exhibit a moderate composition dependence (Fig.~S3 in SI). Subtle changes in the absorption-edge onset, white-line intensity, and higher-energy multiple-scattering features suggest modest modifications of the unoccupied Ga $4p$ states. No pronounced shift of the absorption edge or qualitative change in the overall spectral line shape is observed across the investigated composition range. The local environment around Ga therefore appears less strongly affected by Sc incorporation than the environment around Sc. This contrast between the Sc and Ga absorption edges provides evidence for a site-dependent structural response, with a larger local rearrangement associated with the Sc sites.

\begin{figure}
    \centering
    \includegraphics[width=1\linewidth]{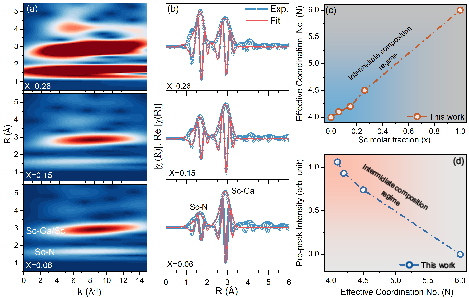}
\caption{(a) Wavelet-transform (WT) maps of the EXAFS signal in combined $k$-$R$ space for Sc concentrations $x=0.06$, $0.15$, and $0.26$. (b) Magnitude of the Fourier-transformed EXAFS spectra (symbols) together with the corresponding best-fit models (solid lines). (c) Effective Sc-N coordination number as a function of Sc molar fraction. (d) Correlation between the normalized Sc $K$-edge pre-edge intensity and the effective Sc-N coordination number.}
\label{fig3:EXAFS}
\end{figure}

The local structural evolution is further examined using Sc $K$-edge EXAFS and wavelet-transform (WT) analysis (Fig.~\ref{fig3:EXAFS}). For $x=0.06$, the first-shell (Sc-N) shows a relatively localized feature in the WT map (Fig.~\ref{fig3:EXAFS}(a)). With increasing Sc concentration, the first-shell contribution broadens along the radial direction, indicating an increasing distribution of Sc-N bond lengths and local structural disorder~\cite{ZHU2026}. The higher-shell Sc-Ga/Sc contributions remain observable across the investigated composition range, indicating that medium-range structural correlations are retained. In contrast to highly Sc-alloyed AlScN, where strong static disorder can substantially attenuate higher-shell EXAFS contributions~\cite{ZHU2026,Knoll2014JPCM}, the persistence of these features in the present samples indicates that the local distortions occur within a wurtzite framework. The EXAFS data do not show evidence for a separate ScN-like local environment or pronounced phase segregation within the investigated composition range.

The Fourier-transformed EXAFS spectra and corresponding fits are shown in Fig.~\ref{fig3:EXAFS}(b). The dominant Sc-N contribution broadens and shifts toward higher $R$ with increasing Sc concentration, consistent with an increase in the average Sc-N bond length and a broader distribution of local bond lengths. The quantitative fitting parameters are summarized in Table~\ref{Table1:EXAFS}. The average Sc-N bond length increases from 2.045(7)~\AA~ at $x=0.06$ to 2.081(8)~\AA~ at $x=0.26$, while the fitted effective Sc-N coordination number increases from 4.1(4) to 4.5(2) (Fig.~\ref{fig3:EXAFS}(c)). Thus the simultaneous increase in the average bond length and fitted coordination number is consistent with a progressive departure from an ideal fourfold tetrahedral environment toward locally distorted configurations with increased effective coordination. Given that the fitted coordination numbers represent effective values within the EXAFS model, they should not be interpreted as evidence for a discrete transformation to sixfold coordination.

Further the Sc-Ga/Sc second shell correlations remain observable throughout the alloy series but become progressively broader with increasing Sc concentration, indicating a wider distribution of local distances around the Sc-centered environment. This evolution is reflected in the fitted Debye-Waller ($\sigma^2$) factor, which increases from 0.0071(1)~\AA$^2$ at $x=0.06$ to 0.014(3)~\AA$^2$ at $x=0.26$. The increase in the $\sigma^2$ is consistent with enhanced static and/or configurational disorder in the local environment. At the same time, the persistence of second-shell correlations indicates that the Sc-induced distortion persist structural ordering. The resulting combination of increased local disorder and retained higher-shell correlations is consistent with a structurally accommodated local distortion rather than an abrupt loss of the wurtzite framework~\cite{Calderon2023Science,Lee2024SciAdv}. 

A direct correlation between the local coordination and the pre-edge intensity can be seen as shown in Fig.~\ref{fig3:EXAFS}(d) where the normalized Sc $K$-edge pre-edge intensity decreases systematically with increasing effective Sc-N coordination number. The inverse relationship between the normalized pre-edge intensity and effective Sc-N coordination provides a quantitative correlation between the local coordination environment and the XANES response. Because the pre-edge intensity is sensitive to local non-centrosymmetry, the observed trend is consistent with a progressive reduction in local asymmetry as the effective coordination increases. Therefore the reduction in local asymmetry is associated with the progressive departure from ideal fourfold coordination. The experimental results therefore provide direct evidence for the composition-dependent evolution of the local Sc environment, while the
microscopic origin and energetic consequences of this evolution are addressed through the first-principles calculations below.

\begin{table*}
\centering
\caption{Best-fit Sc $K$-edge EXAFS parameters for \sgn~ thin films. Fits were performed in $R$-space over the range 1.2-3.7~\AA\ using $k^3$ weighting. Uncertainties represent one standard deviation. Here, x, N, R and $\sigma^2$ are Sc molar fraction, coordination number, bond length and debye-waller factor, respectively.}
\label{tab:exafs_formatted}
\begin{tabular*}{\textwidth}{@{\extracolsep{\fill}}cccccccc@{}}
\toprule
        & \multicolumn{3}{c}{\textbf{First Shell (Sc-N)}} & \multicolumn{3}{c}{\textbf{Second Shell (Sc-Ga)}} & \\
        \cmidrule(lr){2-4} \cmidrule(lr){5-7}
        $x$ & $N$ & $R$ (\AA) & $\sigma^2$ (\AA$^2$) & $N$ & $R$ (\AA) & $\sigma^2$ (\AA$^2$) & $R$-factor \\
        \midrule
        0.06 & 4.1 (4) & 2.045 (7) & 0.0041 (3) & 12.0 (1) & 3.235 (2)     & 0.0071 (1) & 0.027 \\
        0.15 & 4.2 (1) & 2.069 (5) & 0.0038 (1) & 12.1 (5) & 3.251 (2)     & 0.0075 (7) & 0.018 \\
        0.26 & 4.5 (2) & 2.081 (8) & 0.004\hphantom{0} (2)  & 9\hphantom{.0} (2)    & 3.260 (4) & 0.014\hphantom{0} (3)  & 0.086 \\
        \bottomrule
\end{tabular*}
\label{Table1:EXAFS}
\end{table*}

\FloatBarrier

\section{Theoretical insights into coordination-driven structural softening}

To establish the microscopic origin of the experimentally observed structural response, we performed first-principles calculations using special quasi-random structures (SQS) spanning the Sc concentration range $x=0.05$-$0.33$. A representative relaxed supercell is shown in Fig.~\ref{fig4:Fig4DFT}(a), while the complete set of relaxed structures is provided in Fig.~S4 of the SI. The calculations reveal that the structural response is strongly site selective. In ideal wurtzite nitrides, the internal structural parameter is close to $u\approx0.375$, with an ideal axial ratio of $c/a\approx1.633$, corresponding to a nearly tetrahedral local environment \cite{Tasnadi2010PRL,Tholander2013PRB}. With increasing Sc concentration, the composition-averaged $u_{\mathrm{supercell}}$ increases from approximately 0.380 to 0.400, while the axial ratio decreases from 1.623 to 1.598 (Fig.~\ref{fig4:Fig4DFT}(b)). Importantly, the average structural evolution is dominated by the local environment surrounding Sc. The Sc-centered parameter $u_{\mathrm{Sc}}$ increases substantially faster than $u_{\mathrm{Ga}}$, reaching values above 0.41 at intermediate compositions, whereas the Ga-centered environment remains comparatively close to the ideal wurtzite configuration. This site-selective response is consistent with the experimental XANES and EXAFS results, which show a progressive modification of the Sc-centered coordination environment while the overall wurtzite structure is retained.

The composition-dependent structural evolution can be further understood from the corresponding energy landscapes. Figure~\ref{fig4:Fig4DFT}(c) shows a representative potential-energy surface for Sc$_{0.05}$Ga$_{0.95}$N as a function of unit-cell volume and axial ratio, while the complete composition-dependent landscapes are provided in Fig.~S5. The energy surface retains a single wurtzite-derived minimum over the investigated composition range, with no indication of a competing lower-energy minimum. With increasing Sc concentration, the equilibrium structure shifts toward larger volume and lower $c/a$, while the energy surface becomes progressively softer with respect to changes in the axial ratio. Such broadening of the structural energy landscape is consistent with the established picture of Sc-induced lattice softening in wurtzite nitrides \cite{Tasnadi2010PRL,Tholander2013PRB,Paillard2026}. Rather than requiring a transformation to a different crystal structure, the alloy can therefore accommodate an increasingly broad range of locally distorted, wurtzite-derived configurations. In ScGaN, the calculated energy landscape provides a theoretical counterpart to the experimentally observed evolution of the Sc-centered coordination environment.

The energetic consequences of this structural flexibility are reflected in the calculated polarization-switching pathway. Previous theoretical studies have shown that polarization reversal in Sc-containing wurtzite nitrides can proceed through local atomic rearrangements within a wurtzite-derived framework rather than requiring a fully developed competing crystal structure \cite{2019_Fichtner,Ye2025NatCommun,Paillard2026,Zheng2026,Naudin2026}. Using climbing-image nudged elastic band (CI-NEB) calculations, we evaluated the intrinsic switching barrier as a function of Sc concentration. As shown in  0ig.~\ref{fig4:Fig4DFT}(f), the calculated magnitude of the
$P_{\mathrm{sp}}$ decreases progressively with Sc incorporation, from 1.338~C/m$^2$ for GaN to 1.184~C/m$^2$ at $x=0.333$, following an approximately linear dependence. The reduction in the intrinsic barrier accompanies the progressive modification of the Sc-centered bonding environment and the broadening of the structural energy landscape. Within the present theoretical model, the increasing flexibility of the local coordination environment allows intermediate configurations along the polarization-reversal pathway to be accommodated at lower energetic cost. We note that the CI-NEB calculations were performed at fixed cell shape and volume; consequently, the absolute barriers should be regarded as upper-bound estimates, whereas the composition-dependent trend provides the more robust quantity for comparison \cite{Lee2024SciAdv}.

The electronic structure provides further insight into the origin of this local structural flexibility. The orbital-resolved projected density of states (PDOS) in Fig.~\ref{fig4:Fig4DFT}(e) shows that the valence-band states are dominated by N~$2p$ orbitals, whereas the conduction-band region contains substantial Sc~$3d$ character hybridized with N~$2p$ and Ga-derived states. The inset resolves the Sc~$3d$ contribution into individual orbitals and reveals a pronounced orbital dependence. In particular, the $d_{xy}$ and $d_{x^2-y^2}$ states are comparatively localized, whereas the out-of-plane $d_{z^2}$ contribution is distributed over a broader energy range. The orbital anisotropy is consistent with the directional structural response observed experimentally and suggests that Sc-derived electronic states may contribute differently to distortions along the polar and in-plane directions.

\section{Functional consequences: polarization and electromechanical response}

\begin{figure}
\centering
\includegraphics[width=1\linewidth]{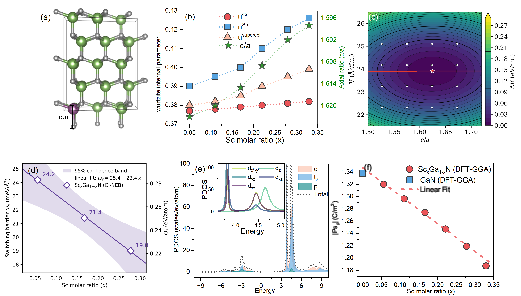}
\caption{(a) Representative relaxed SQS supercell of Sc$_{0.05}$Ga$_{0.95}$N. (b) Composition dependence of the site-resolved internal parameters around Ga ($u_{\mathrm{Ga}}$) and Sc ($u_{\mathrm{Sc}}$), the composition-averaged internal parameter ($u_{\mathrm{supercell}}$), and the axial ratio ($c/a$). (c) Representative potential-energy landscape of Sc$_{0.05}$Ga$_{0.95}$N as a function of unit-cell volume ($V$) and axial ratio ($c/a$). (d) Composition dependence of the intrinsic polarization-switching barrier ($\omega_s$) obtained from climbing-image nudged elastic band calculations. The shaded region represents the 95\% confidence band of the linear fit. (e) Orbital-resolved projected density of states of Sc$_{0.05}$Ga$_{0.95}$N. The inset highlights the Sc $3d$ orbital contributions near the conduction-band edge. (f) Calculated magnitude of the spontaneous polarization, $|P_{\mathrm{sp}}|$, as a function of Sc concentration. The dashed line represents a linear fit.}
\label{fig4:Fig4DFT}
\end{figure}

The coordination-driven structural softening identified above has important consequences for the intrinsic polar and electromechanical properties of {\sgn}. We therefore examine the composition dependence of the spontaneous polarization (P$_{sp}$), piezoelectric coefficients, and elastic response. The P$_{sp}$ was obtained using a dense Berry-phase path and independently cross-checked against a point-charge polarization reference to resolve the polarization branch, as described in the section D in SI. The evolution of the P$_{sp}$ provides a measure of the changing equilibrium polar distortion. As shown in Fig.~\ref{fig4:Fig4DFT}(f), the calculated magnitude of the P$_{sp}$ decreases progressively with Sc incorporation, from 1.338~C/m$^2$ for GaN to 1.184~C/m$^2$ at $x=0.333$, follow an approximately linear dependence. This reduction is consistent with the decrease in the axial ratio and the evolution of the internal structural parameters, indicating that Sc incorporation reduces the magnitude of the equilibrium polar distortion while simultaneously broadening the structural energy landscape. The calculated GaN endpoint is also consistent with established first-principles values, supporting the reliability of the polarization protocol \cite{Bernardini1997,Zoroddu2001}.

To quantify the electromechanical consequences of the structural softening, we calculated the piezoelectric stress (e$_{ij}$) and strain tensors (d$_{ij}$) together with the elastic constants (C$_{ij}$) for $x=0$-$0.333$. The complete tensor components are provided in Fig.~S7, with the corresponding computational details given in Section D of the SI. The longitudinal piezoelectric stress coefficient $e_{33}$ increases monotonically from 0.82~C/m$^2$ for GaN to 1.66~C/m$^2$ at $x=0.333$, corresponding to an approximately two-fold enhancement. In parallel, the longitudinal elastic stiffness $C_{33}$ decreases from 380 to 227~GPa, corresponding to a reduction of approximately 40\%. The calculated GaN endpoint values of $e_{33}=0.82$~C/m$^2$ and $C_{33}=380$~GPa are in reasonable agreement with previously reported first-principles values of 0.66~C/m$^2$ and 354~GPa, respectively, while the calculated $d_{33}=3.0$~pC/N is consistent with previously reported values for GaN \cite{Bernardini1997,Zoroddu2001}. This agreement provides a useful validation of the computational approach used to evaluate the electromechanical response. The stronger composition dependence of the piezoelectric strain response can be understood from the combined evolution of the piezoelectric and elastic tensors. The relationship between the piezoelectric stress and strain coefficients is given by
\begin{equation}
d_{ij}=\sum_k e_{ik}S_{kj},
\end{equation}
where $S=C^{-1}$ is the elastic compliance tensor. The calculated $d_{33}$ increases from 2.99~pC/N for GaN to 12.46~pC/N at $x=0.333$, corresponding to an approximately 4.2-fold enhancement. Thus, the increase in $d_{33}$ reflects the combined effects of enhanced piezoelectric coupling and increased elastic compliance associated with the structural softening. In particular, the pronounced reduction in $C_{33}$ indicates a substantial decrease in longitudinal stiffness and contributes to the enhanced compliance along the polar direction. Consequently, the relative enhancement of $d_{33}$ is considerably larger than that of $e_{33}$, demonstrating that elastic softening makes an important contribution to the calculated electromechanical response.

The calculated trends are consistent with previous theoretical studies of Sc-alloyed GaN and related Sc-containing wurtzite nitrides, in which enhanced piezoelectric response has been associated with increasing $e_{33}$ and decreasing $C_{33}$ \cite{Paillard2026,Tasnadi2010PRL}. At $x=0.333$, the calculated values of $e_{33}=1.66$~C/m$^2$ and $d_{33}=12.5$~pC/N remain below the larger enhancements predicted at higher Sc concentrations, consistent with the moderate alloying range considered here \cite{Wang_2021,Yang_2025}. The full composition dependence of the remaining piezoelectric and elastic tensor components is provided in Fig.~S7.

It is important to distinguish the intrinsic material response calculated here from the macroscopic response obtained from electrical ferroelectric measurements. Electrical measurements of ScGaN have also shown that leakage can contribute
to the measured $P$-$E$ response, motivating the use of PUND measurements separate switching-related and nonswitching contributions~\cite{Yang_2025,Uehara_2022}. These extrinsic factors make direct comparison between calculated intrinsic properties and experimentally extracted macroscopic polarization or coercive-field values nontrivial. The present calculations therefore provide a complementary description of the intrinsic response of the ScGaN lattice, rather than a direct reproduction of device-level ferroelectric measurements. Within this framework, the systematic reduction of the calculated switching barrier and the simultaneous enhancement of the electromechanical response with increasing Sc concentration provide insight into how the coordination-mediated structural softening influences the intrinsic functional response of the alloy.

Taken together, the experimental measurements and first-principles calculations provide a consistent picture of the composition-dependent response of {\sgn}. Sc incorporation preferentially modifies the local Sc-centered environment while the surrounding Ga-centered wurtzite framework remains comparatively preserved. This local coordination evolution is accompanied by broadening of the structural energy landscape, a reduction in the calculated intrinsic polarization-switching barrier, pronounced longitudinal elastic softening, and an enhanced piezoelectric strain response, while the magnitude of the spontaneous polarization decreases progressively with Sc incorporation. These results indicate that the coordination flexibility introduced by Sc provides a common structural basis for the calculated energetic and electromechanical responses of ScGaN. The combined experimental and theoretical results therefore extend the coordination-mediated picture established for Sc-containing wurtzite nitrides to ScGaN and highlight local coordination as an important microscopic parameter governing its structural and functional response.

\section{Conclusion}

In summary, we have investigated the structural evolution and local bonding response of Sc-alloyed wurtzite GaN by combining XRD, XANES, EXAFS, and first-principles calculations. The diffraction measurements reveal a pronounced anisotropic lattice response, characterized by a stronger increase in the in-plane lattice parameter and a progressive reduction in the $c/a$ ratio with increasing Sc concentration, while the long-range wurtzite structure is retained. Sc $K$-edge XANES and EXAFS further show a systematic modification of the local Sc environment, manifested by a reduction in pre-edge intensity, increasing Sc-N bond length, increasing effective coordination, and enhanced local structural disorder. The first-principles calculations further provide a microscopic interpretation of these experimental observations, revealing pronounced distortions around Sc and a progressively softer energy landscape with respect to polar structural distortions. This structural flexibility is accompanied by a reduction of the calculated intrinsic polarization-switching barrier from 24.2 to 19.0~meV/\AA$^3$ over the investigated composition range. The spontaneous polarization also decreases from 1.338~C/m$^2$ for GaN to 1.184~C/m$^2$ at $x=0.333$, consistent with the progressive flattening of the wurtzite structure. The calculated electromechanical response further demonstrates the functional consequences of this structural evolution. The longitudinal piezoelectric coefficient $e_{33}$ increases from 0.82 to 1.66~C/m$^2$, while $C_{33}$ decreases from 380 to 227~GPa with increasing Sc content. As a consequence of the simultaneous enhancement of piezoelectric coupling and elastic compliance, the calculated $d_{33}$ increases from 2.99 to 12.46~pC/N. These results indicate that Sc-induced local structural flexibility is accompanied by substantial changes in the intrinsic electromechanical response of \sgn{}.

Overall, this work extends the established picture of Sc-induced structural softening in wurtzite nitrides to \sgn~ by experimentally resolving the evolution of the Sc-centered local environment and, through first-principles calculations, relating it to structural energetics, polarization, switching barriers, and electromechanical response. The results highlight local coordination as an important microscopic descriptor for understanding and tuning the coupled structural and functional properties of metastable wurtzite nitrides.

\bibliography{apssamp}

\section{\label{Sec:level1}Methods}
\subsection{Thin Film Growth}
\sgn~ films were grown on 10$\times$10\,mm$^2$ c-plane sapphire substrates by direct current (DC) reactive magnetron sputter epitaxy at a base pressure below 8 $\times$ 10$^{-9}$ Torr. To ensure chamber cleanliness and reproducibility, the substrates were out-gassed at 950$^\circ$C for 15 min, then cooled to the growth temperature 400$^\circ$C before deposition. The deposition was performed at a working pressure of 3 mTorr with an N$_{2}$/Ar partial-pressure ratio of 3:7. The substrate potential was kept floating and was rotated at 17 rpm to ensure uniformity. To vary the relative Ga and Sc concentrations, the Ga and Sc target powers were varied as follows: 60, 59, 52, 45, 0 and 0, 68, 143, 219, 135, 250 W, respectively. More detailed about the growth parameters are provided elsewhere \cite{Hsiao2026APL}.  

\subsection{Experimental Characterization}
The crystal structure of samples was analyzed by $\omega$-2$\theta$  XRD using a PANalytical diffractometer with a Cu-K$\alpha$, $\lambda$ = 1.5406 Å source. The symmetric \(\omega-2\theta\) scans were performed in Bragg–Brentano geometry, while the asymmetric \((10\bar11)\) reflection was measured using an appropriate tilted geometry at $\chi \approx$ 61.9$^\circ$. The composition of the films was measured by a combination of Rutherford backscattering spectrometry (RBS) and time-of-flight elastic recoil detection analysis (ToF-ERDA)~\cite{Strom_2022}. 2 MeV He$^+$ primary ion beam was incident at 5$^\circ$ to the sample normal, and the backscattered particles were detected at an angle of 170$^\circ$. For ToF-ERDA measurements, a 36 MeV iodine ($^127$I$^{8+}$) primary beam was incident at an angle of 67.5$^\circ$ with respect to the sample surface normal. The light element composition was extracted from elemental depth profile using Potku code~\cite{ARSTILA201434}. To investigate the local and electronic structure, X-ray absorption spectroscopy (XAS) measurements were performed at room temperature (RT) in fluorescence mode at Balder beamline at MAX IV synchrotron radiation source. The XAS data were normalized using pre and post edge corrections and the EXAFS spectra were analyzed using the FEFFIT implementation within the Larch package \cite{Newville2013}. Theoretical scattering paths were generated using FEFF calculations based on the wurtzite Sc-substituted GaN structural model \cite{Rehr2000,Ankudinov1998}. The experimental $\chi(k)$ data were weighted by $k^3$ and Fourier transformed over the range 2.0-14 Å$^\text{-1}$ using a Kaiser-Bessel window function ($dk = 3$ Å$^\text{-1}$). The fitting was performed in $R$-space over the interval 1.1-3.7 Å, encompassing the first Sc-N coordination shell and the second Sc-Ga shell. 

\subsection{Computational Details}

First-principles calculations were performed within the framework of density functional theory (DFT) as implemented in the Quantum ESPRESSO package \cite{Giannozzi2009JPCM,Giannozzi2017JPCM,Giannozzi2020JCP,QuantumEspressoURL}. The exchange-correlation potential was treated using the generalized gradient approximation (GGA) in the Perdew-Burke-Ernzerhof (PBE) formulation. Structural relaxations of $\mathrm{Sc}_x\mathrm{Ga}_{1-x}\mathrm{N}$ alloys were carried out using periodically repeated $3\times3\times2$ wurtzite supercells containing 72 atoms (36 cations and 36 anions). Substitutional Sc atoms were distributed over Ga sites using the special quasi-random structures (SQS) approach \cite{vandeWalle2013Calphad}, generated with the ATAT toolkit \cite{vandeWalle2002Calphad,vandeWalle2009Calphad,Lebeda2026SimplySQS,AtatSqsURL}, yielding alloy compositions from $x\approx0.056$ (2 Sc atoms) to $x\approx0.333$ (12 Sc atoms). Brillouin-zone integrations for structural relaxations employed a $2\times2\times2$ Monkhorst-Pack $k$-point mesh, while projected density of states (PDOS) calculations were performed using a denser $15\times15\times15$ mesh. Both lattice parameters and internal atomic coordinates were fully relaxed until the residual forces on all atoms were below $1\times10^{-4}$~eV/\AA.

The evolution of the structural parameters was analyzed through the composition dependence of the axial ratio $c/a$ and the internal structural parameter $u$, defined from the relative displacement of the cation and anion sublattices along the polar $c$ axis. Site-resolved internal parameters ($u_{\mathrm{Sc}}$ and $u_{\mathrm{Ga}}$) were extracted directly from the relaxed supercell geometries to quantify the local structural distortion around Sc and Ga atoms separately.

The structural energetics were further investigated by mapping the total energy as a function of unit-cell volume $V$ and axial ratio $c/a$. For each composition, a $5\times5$ grid spanning $\pm8\%$ in volume and $\pm6\%$ in $c/a$ around the equilibrium structure was constructed. All internal atomic coordinates were fully relaxed at each grid point, and the resulting energy surfaces were fitted using a second-order polynomial expansion to obtain the composition-dependent potential-energy landscapes.

Polarization switching barriers were calculated using the climbing-image nudged elastic band (CI-NEB) method \cite{Henkelman2000} as implemented in the \texttt{neb.x} module of Quantum ESPRESSO. The minimum-energy path connecting the two equivalent polar wurtzite states (${\mathrm{wz}^{+}}$ and ${\mathrm{wz}^{-}}$) was discretized into 21 images generated by linear interpolation, with one climbing image constrained to converge to the saddle point. The endpoint structures were fully relaxed prior to the NEB calculations. Total energies and forces were computed using the PBE functional with Grimme D3 dispersion corrections \cite{Grimme2010}, a plane-wave cutoff of 70 Ry (560 Ry for the charge density), cold smearing of 0.01 Ry, and a $2\times2\times2$ Monkhorst-Pack $k$-point mesh. The switching barrier $\mathrm{\omega_s}$ is reported as a volumetric energy density (meV/\AA$^{3}$), obtained by normalizing the activation energy by the relaxed formula-unit volume, following previous studies of ferroelectric wurtzite nitrides \cite{Lee2024SciAdv}. We note that the present calculations employ a fixed-cell CI-NEB approach in which the lattice vectors are held constant along the switching path. Consequently, the calculated barriers represent upper bounds to the intrinsic switching barrier, whereas the composition dependence and overall trend of $\mathrm{\omega_s}$ remain robust. \cite{Sheppard2012,Zheng2026}

To further quantify the electromechanical response associated with the structural evolution of \sgn, first-principles calculations of the piezoelectric stress and strain tensors, elastic constants, and spontaneous polarization were performed using the same PBE+D3 framework.
A series of $3\times3\times2$ wurtzite supercells containing 72 atoms and 36 cation sites was considered, corresponding to Sc concentrations of $x=0$, 0.056, 0.111, 0.167, 0.222, 0.278, and 0.333. More details are given in the piezoelectric and elastic response section in SI.

Sc $K$-edge XANES spectra were calculated using the finite-difference method implemented in the \textsc{FDMNES} code \cite{Joly2001,Joly2009}. Self-consistent full-potential calculations including dipole and quadrupole transitions were performed using clusters containing approximately 180 atoms surrounding the absorbing Sc atom. No symmetry constraints ($C\mathrm{_{1}}$ symmetry) were imposed in order to capture the local structural distortions around Sc. PDOS calculations were used to analyze the orbital origin of the pre-edge features, while theoretical spectra were broadened to account for the core-hole lifetime and instrumental resolution.

\section{Data availability}
The data that support the findings of this study are available in the
article and its Supplementary Information. 

\section{Competing interests}
The authors declare no competing interests.

\section{Acknowledgment}
The authors gratefully acknowledge Daniel Primetzhofer and Mauricio Sortica for RBS and ToF-ERDA measurements at Tandem Laboratory, Uppsala University which is financed by the Swedish Research Council VR-RFI under contract number 2019-00191. Issa Nseir is acknowledged for help in sample preparation. Also Research conducted at MAX IV, a Swedish national user facility, is supported by the Swedish Research Council
under Contract No. 2018–07152, the Swedish Governmental Agency for Innovation Systems under Contract No. 2018–04969, and Formas under Contract No. 2019–02496. Konstantin Klementiev from Balder beamline, MAX IV, Lund, Sweden, is acknowledged for the help during the XAS measurements. This work was supported by the Olle Engkvists Stiftelse (238-0091 (C.-L.H.), 227-0244 (C.-L.H.)); Swedish Research Council (2025-04747 (R.-H.H.), 2025-03705 (M.M.), 2021-03826 (P.E.), 2025-03680 (P. E.)); Carl Tryggers Stiftelse (CTS 24:3577 (C.-L.H.), CTS23:2746 (C.-L.H.), CTS 25:3972 (M.M.), CTS23:2746 (M.M.)); Swedish Government Strategic Research Area in Materials Science on Functional Materials (2009 00971) and Knut and Alice Wallenberg Foundation (KAW-2020.0196, WISE-IP02-PD07 (C.-L.H.)). The computations were enabled by resources provided by the National Academic Infrastructure for Supercomputing in Sweden (NAISS) at PDC, KTH Royal Institute of Technology (Dardel), under project nr. NAISS 2026-3-216, partly funded by the Swedish Research Council (2022-06725 (M.M.)).

\section*{Author contributions}

S. K. and C-L. H. conceived the project. S. K., M. M. and C-L. H. carried out the X-ray absorption measurements and S. K. analyzed the XANES and EXAFS data. G. K. G. and E. F. performed the first-principles calculations, including structural relaxations, electronic structure, energy-landscape, and switching-barrier calculations. S. J. analyzed RBS and ERDA data. S. K. and C-L. H. interpreted the results with input from R. S., P. S., R-H. H., J. B., and P. E. S. K. wrote the original draft of the manuscript with contributions from all authors. All authors discussed the results and approved the final version of the manuscript.

\end{document}